# Challenges of capturing engagement on Facebook for Altmetrics

Asura Enkhbayar[*], Juan Pablo Alperin[*]

[*]aenkhbay@sfu.ca; juan@alperin.ca
Scholarly Communications Lab, Simon Fraser University, 515 West Hastings Street
Vancouver, BC, V6B 5K3 (Canada)

**Introduction**
With almost twice the number of active monthly users than the population of China and India, Facebook is by far the biggest social network (Facebook, 2018). By comparison, the number of active users on Twitter is one sixth the size (Twitter, 2018). Naturally, users on both platforms include academics of all kinds. It is therefore surprising that several studies have been consistently reporting differences between the coverage of papers on Twitter and Facebook. Two older studies reported 27.7% and 20% coverage for Twitter in contrast to 11.3% and 2.9% respectively (Thelwall et al., 2013; Hammarfelt, 2014). A recent study by Zahedi and Costas (2018) looked into the differences across several data aggregators and once again Twitter repeatedly showed higher coverage of papers on Twitter (57% of articles receive at least one Tweet, while only 16.3% of them are shared on Facebook). To better understand the discrepancy between the popularity of the platform and the relatively low levels of reported use for sharing academic content, we set out to look for additional engagement of research on Facebook by searching for articles through the Facebook Graph API. However, before we could explore the discrepancy, we encountered several fundamental challenges of collecting engagement data from Facebook. These difficulties partially overlap with previously identified challenges of altmetrics (Haustein et al, 2016; Chamberlain, 2013), while others are specific to working with APIs and URLs. This paper presents these challenges along with a first approximation of how pervasive the problems they generate are.

**If there were no challenges**
In an ideal world, collecting engagement about scholarly articles would follow a simple pathway: 1) A document would be identified by a Digital Object Identifier (DOI); 2) Crossref would provide the most recent URL associated with that DOI; 3) the Graph API would be queried with the URL; 4a) Facebook would map this URL to their internal identifier system; and 4b) it would simultaneously return the number of its engagements (see Figure 1).

*Figure 1: The ideal collection process; never encountered in real life.*

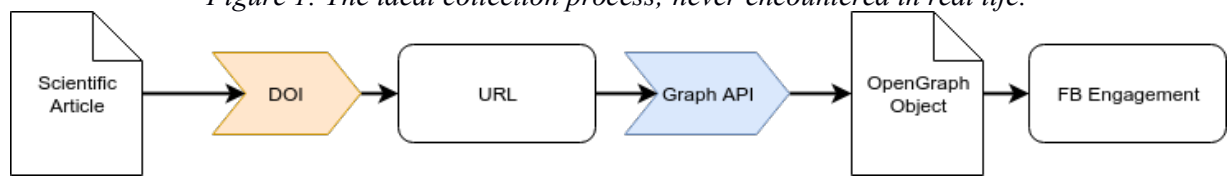

When working with Facebook's Graph API to collect altmetrics, we encountered two main types of challenges, each at different stages of the collection process. Some of these problems are specific to working with Facebook's interface, while others are general problems of collecting data about objects online.

The first challenge in collecting metrics is **mapping an article to the URLs where it resides** (both current URL and previous locations). While information on the web is identified by a URL, in the scholarly realm documents are commonly identified by a DOI. Unfortunately, as Wass (2016) points out in his detailed analysis, the relationship between DOIs and a URLs is a complicated one. In our simplified scenario, there is a one-to-one mapping of DOI to URL, but, in reality, most scholarly articles exist at multiple URLs. The DOI itself, for example, is a URL (e.g., https://doi.org/10.5555/12345678) that, if working correctly, will lead the browser to the article. Moreover, even when everything is as it should be, articles can still reside on other platforms (e.g., Pubmed or self-archived versions on author's homepages). While some services and tools provide ways to discover other versions (e.g., OAI-PMH, Unpaywall, and BASE) often times there is no automatable solution to this problem (e.g., Google Scholar detects self-hosted versions but does not provide an API). However, even if choosing to work from a fixed and knowable set of URLs, there is a second set of challenges.

Given a set of URLs for an article, there is a challenge in understanding **how to aggregate the metrics that are collected for each**. Depending on the structure of a web page, different URLs might actually point to the same final location (e.g., a doi.org URL and the location it resolves to), so it is important to make sure that the metrics collected for each are not double-counted. In other cases, different URLs correspond to the same article at different locations, and therefore those metrics should be added together to arrive at the overall metrics for the document. In the case of Facebook, this is done by mapping each URL to a Facebook Open Graph Object with a unique ID (Ob_ID). Unfortunately, this mapping of multiple URLs back to a single identifier is not straightforward, even when care is taken to follow best practices. Taken together, both sets of challenges mean that any metrics aggregator attempting to use URL-based APIs will need to accept some errors and limitations.

While Figure 1 shows an overly simplified diagram of the data collection process as it might be imagined in an ideal scenario, a more complex figure that takes into account the multiple versions of articles is presented in Figure 2. This figure outlines another idealized—but more accurate—case of the collection process. To understand these challenge areas, we provide examples and explore their prevalence by using the random sample of found articles in the Web of Science used by Piwowar et al. (2017). Their full dataset can be found at [DOI 10.5281/zenodo.1041791] while our code is stored at [DOI 10.5281/zenodo.1317598].

**Challenge Area 1: Mapping articles to URLs**
Anyone trying to collect social media metrics will need to be concerned with identifying where an article resides online. As Chamberlain (2013) observes, such digital provenance is usually provided by URLs or identifiers, but where and how to collect these is not straightforward. This challenge has, to some degree, been documented (Chamberlain, 2013; Wass, 2016; Liu & Adie, 2013). However, although each article lives at multiple URLs and is connected to various identifiers across platforms (as shown in Figure 2), these links are subject to change, disappear, or sometimes simply point to wrong locations. Other times, even the backbone of academic linking—the DOI system—simply fails to work.

*Figure 2: A more complete but still idealized collection process*

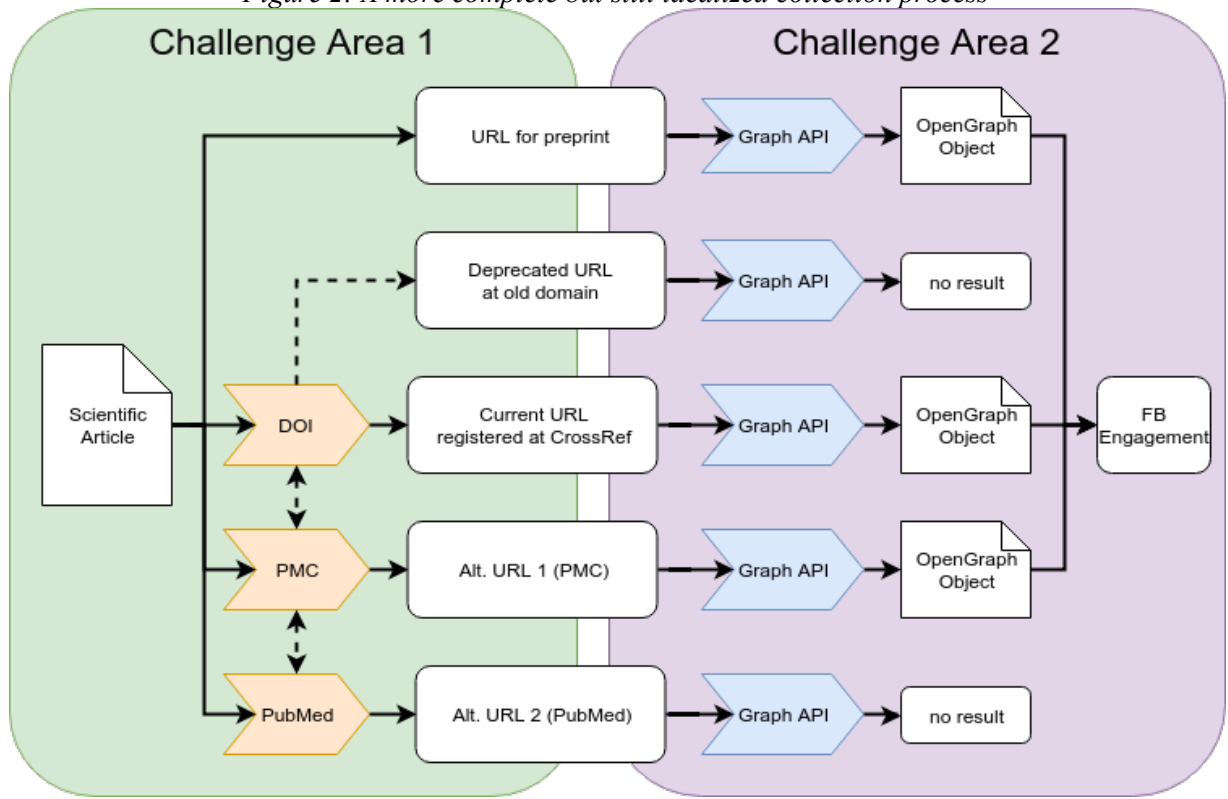

*Problems Case 1 - Identifying the landing page from any given DOI*
Exploring the "extraordinary diversity" (Wass, 2016; n.p.) of how DOIs map to URLs is an endlessly complex task. As an employee of Crossref, Wass is uniquely positioned, both in terms of knowledge and access to data, to explore and the challenges in mapping DOIs to landing pages of articles. With a sample of over 11M DOIs, he outlines the categories that DOI resolutions can fall into, and through a series of experiments, explores the many "nooks and crannies" that complicate the mapping of DOIs to landing pages. We do not repeat his experiments here but attempt to document, given our random set of DOIs, how those working from outside Crossref can use a DOI to identify the URL of an article landing page. In particular, we used our sample set to check whether we could use HTTP requests to resolve URLs at all (Figure 3, document C), and to see whether these pointed to duplicate URLs (Figure 3; documents A and B).

*Figure 3: A first set of problems with DOIs where multiple DOIs are mapped to the same URL(A + B) or are not mapped to any URL at all (C).*

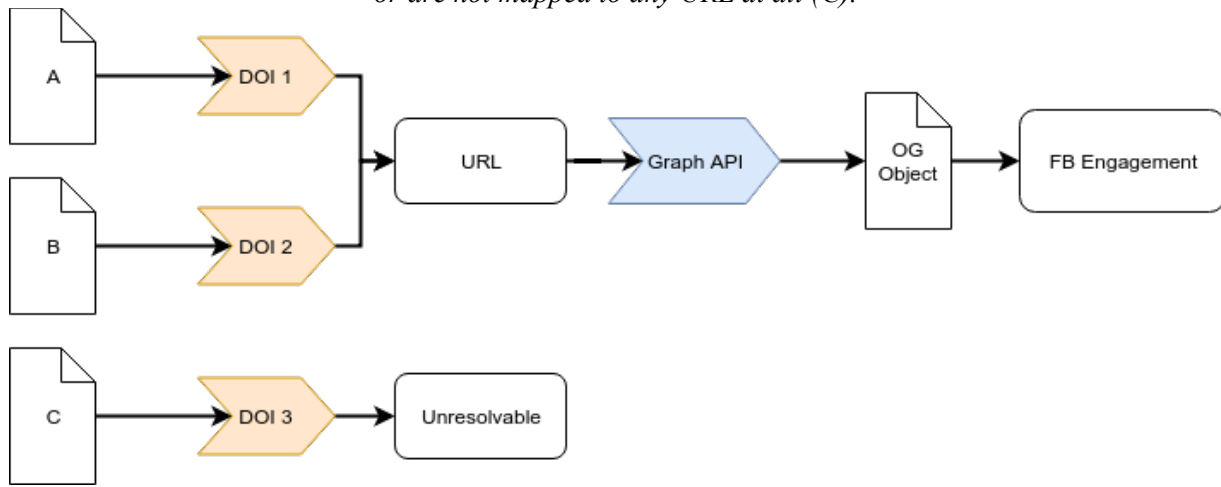

We used HTTP GET requests, with a 5-second timeout, to attempt to resolve the 103,539 DOIs (category 2 from Wass, 2016). Table 1 shows the number of requests that successfully returned a URL (with and without an error code), as well as those that failed, either because the target page required Javascript or cookies, or because of timeouts are shown. This test suggests that tools that rely on HTTP to identify the landing page of an article fail in 11.6% of the cases and need to be reviewed more closely in an additional 5.8% of the time. Additionally, we checked for the scenario where multiple DOIs resolve to the same landing page and found this happened only 68 (0.1%) times out of the 91490 successful resolutions (far less than Wass' estimate of 1%).

*Table 1. Number of successful and problematic attempts to resolve DOIs to URLs from a random set of DOIs from Web of Science*

|  | Number of responses | |
| --- | --- | --- |
| Returned URL successfully | 85,515 | 82.6% |
| Returned URL, with error code* | 5,975 | 5.8% |
| **Total resolved URLs** | 91,490 | 88.36% |
| Failed requests** | 12,049 | 11.6% |
| **Total** | 103,539 | 100% |

*The HTTP GET request returned an error, but still resolved to a URL
**The HTTP GET request was either aborted from server side or timed out after 5s

While we agree with Wass (2016) that "we can find the Landing Page for every DOI … most of the time," the problem of identifying relevant URLs does not end there. From any given landing page there can be a wide range of URLs (e.g., abstract page, full-text page, PDF/download link) that readers arrive at or click before sharing the article on Facebook, which brings us to the second set of challenges.

**Challenge Area 2: Mapping URLs to Open Graph objects**
Even assuming that we have successfully resolved DOIs, found relevant IDs, and determined the relevant URLs for an article, the next challenge poses itself once we engage with Facebook's Graph API and start to query results for these URLs. According to Facebook's

Open Graph protocol ("Graph API," n.d.), many of these URLs should be represented by a single Open Graph Object, with a single Ob_ID, and corresponding canonical URL. This ideal scenario is depicted in Figure 4. Facebook encourages content providers to include meta tags to indicate the canonical URL for equivalent pages ("Best Practices - Sharing" n.d.). Without such meta tags, Facebook relies on a set of heuristics to test the similarity of URLs. Unfortunately, we encountered that this mapping did not always work, even when the appropriate meta tags were present.

*Figure 4: The ideal mapping between an article with multiple URLs and an Open Graph object*

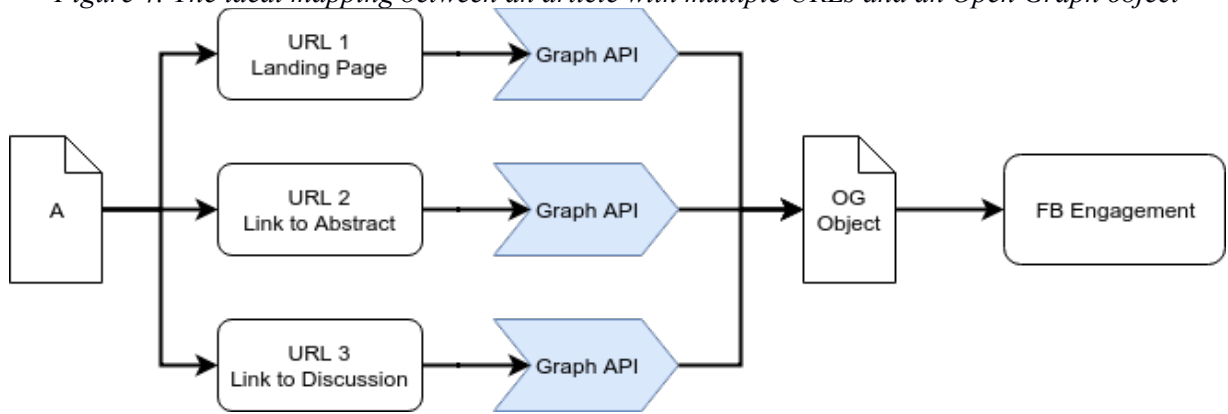

*Problems Case 2 – Equivalent URLs mapped to different OG Objects*
Given that articles can live at multiple URLs, it was important to test the cases when Facebook's Graph API successfully mapped those URLs back to the same Object (and, ideally, returning the same engagement numbers). If a set of URL variants returned the same Graph Object, it would only be necessary to query all the variants. We found this to not be the case, even when using seemingly equivalent URLs. We found the case of mismatches described in Figure 5 querying four URLs for each DOI:

1. the URL where the DOI resolved,
2. the "opposite" protocol URL (http vs https, and vice versa),
3. the currently recommended syntax by Crossref https://doi.org/[doi], and
4. the older syntax http://dx.doi.org/[doi].

*Figure 5: A second problem case where different URLs for the same article are considered a different object by Facebook*

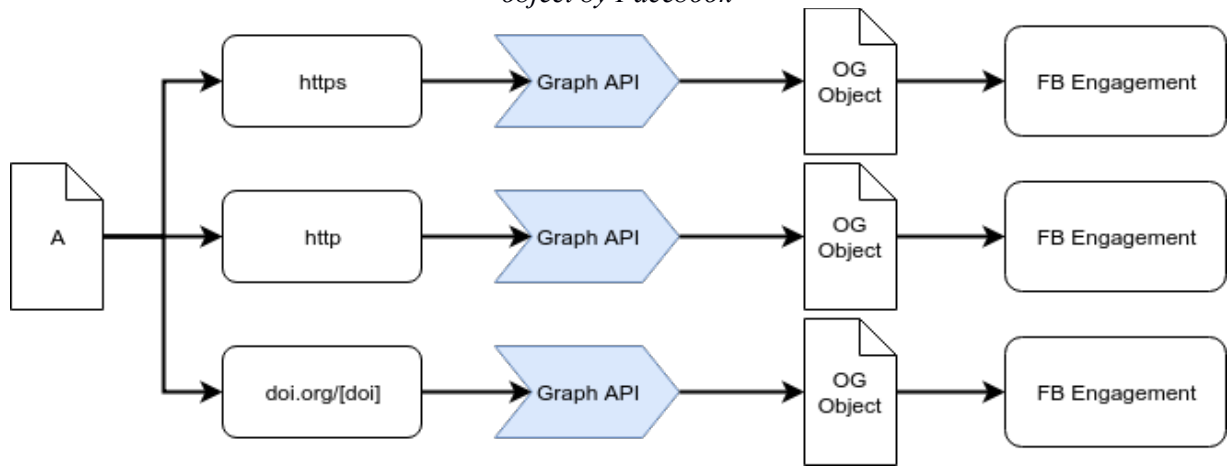

For every URL queried, the Graph API returns two separate entities: the Open Graph Object (with its corresponding Ob_ID) and its engagement. Many queries that return a Graph Object do not report engagements (i.e., Facebook knows or finds out about the URL, but they do not report any engagements for it). We report the coverage (number and percent of articles with Objects and engagements found in the API (Table 3). Of the 91,490 DOIs that we were able to resolve, the coverage of Objects found (i.e., have Ob_ID) ranged from 0.2% to 9.2%, depending on the URL variant. The coverage for articles with engagement ranged from 0.1% to 2.9% across the four variants (Table 3).

*Table 2. Responses from Facebook Graph API for each URL variant*

| Variant | Description | Responses with at least one Ob_ID (N=91490) | | Responses positive engagement (N=91490) | |
|---|---|---|---|---|---|
| 1 | URL where DOI resolved[*] | 8,452 | 9.2% | 1,426 | 1.6% |
| 2 | The "opposite" protocol URL[*] | 13,305 | 14.5% | 2,458 | 2.7% |
| 3 | The current recommended DOI syntax (https://doi.org/[doi]) | 179 | 0.2% | 74 | 0.1% |
| 4 | The older DOI syntax (http://dx.doi.org/[doi]) | 10,124 | 11.1% | 2,612 | 2.9% |
| **All** | **Any of the above variants** | **26,775** | **29.3%** | **5,498** | **6.0%** |

[*]21,871 (23.9%) DOIs resolved to http and 69,619 (76.1%) resolved to https

Already, we can see that which URL is queried matters, despite the fact that each is expected to point to the same landing page. These differences lead us to the next set of problems: deciding how to combine the engagement numbers obtained for each variant. To explore this, we also compared the Ob_ID we obtained for each query with the corresponding engagement numbers and found that there are differences between these seemingly equivalent URLs for almost all DOIs (Table 4).

Of the 5,498 DOIs for which we found at least one engagement, all but 6 were mapped to either different Ob_IDs or returned different engagement numbers (including instances where one URL returned an Ob_ID while another for the same DOI did not). By far the most common case was for articles for which only one variant returned engagement numbers (3,687, 67.1%). There was a small (106; 1.9%)–but perplexing–group of articles for which the API did not return an Ob_ID, but did return engagement numbers. For the remaining cases, at least two of the URL variants return an Ob_ID. These cases can be divided into three overlapping categories: those that have at least two non-matching IDs, those that returned at least two matching ID with matching engagement numbers, and those that returned at least two matching IDs with different engagement numbers for each. Of these, the latter category is the most problematic, because it is impossible to determine which engagement numbers are correct, or if they should be combined. In contrast, mismatched IDs suggest Facebook considers the URLs different and so the engagement numbers could be added.

*Table 3. Number of cases for each scenario of how the Facebook API can respond to the four URL variants*

| Case description | Number | Not matching IDs | Matching ID (matching shares) | Matching IDs (not matching engagements) |
| --- | --- | --- | --- | --- |
| No variant returned an Ob_ID* | 106 | - | - | - |
| One variant returns an Ob_ID | 3,687 | - | - | - |
| Two variants return an Ob_ID | 1,535 | 769 | 620 | 146 |
| Three variants return an Ob_ID** | 161 | 131 | 99 | 43 |
| Four variants return an Ob_ID** | 9 | 8 | 6 | 3 |
| **Total** | **5,498** | **908** | **725** | **192** |

* Although it should not be possible to have engagements without having an Ob_ID, we found some instances where this was the case.
** In some cases, two or three of the Ob_IDs matched, but one or two did not; such cases are counted under all of the appropriate columns.

Overall, these number might imply that when we restrict our queries to four simple and equivalent URLs that all resolve to the same final page, we could resolve the engagement numbers (by either choosing one or combining several responses) in all but 192 (3.5%) of the cases for which there was at least one response with engagements. Unfortunately, there are more problematic cases.

*Problems Case 3 – Different articles are mapped onto the same Graph Object*
Although we might conclude that it is safe to sum the engagements for all the different Graph Objects we identified for a given article, we sometimes find that the same Graph Object has been linked to multiple articles. This case is analogous to Problem Case 1, except that instead of multiple DOIs being mapped to the same URL, we have multiple URLs being mapped to the same Graph Object. This scenario is depicted in Figure 6.

*Figure 6: A third problem case where the URLs for different articles are considered the sameobject by Facebook*

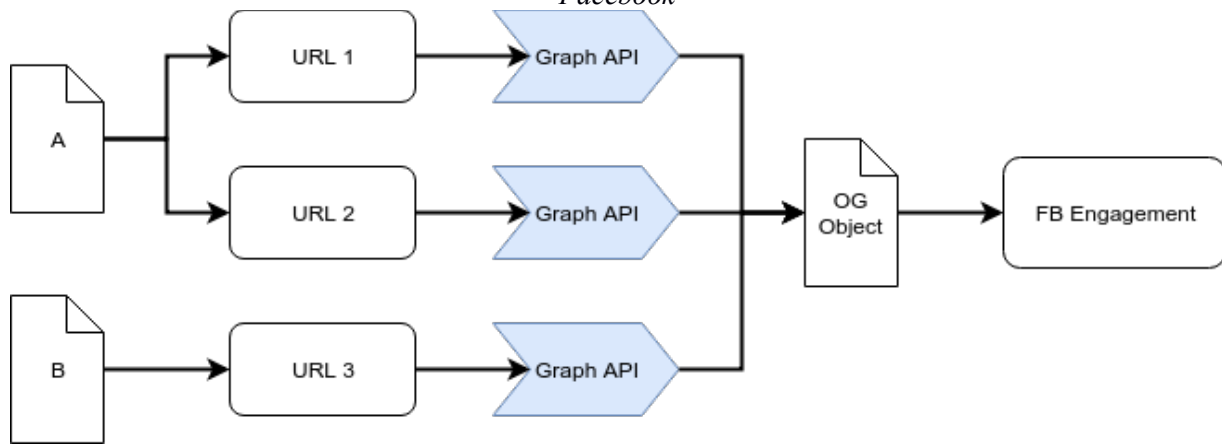

We identified a total of 66 Ob_IDs (0.2% of 28711) that were linked to multiple DOIs came up for more than one DOI, with one ID coming up for 184 articles. In total, these 66 Ob_IDs were linked to 507 articles, including 482 of the 5,498 (8.8%) with positive engagements. It is important to note that this case is more likely to appear when looking at larger datasets from same domains which are more likely to get confused as equivalents.

Taken together, problems cases 2 and 3 identified a combined total of 648 articles (11.8% of those 5,498 with positive engagements). When we add those to the 12,049 that we were unable to resolve, our limited exploration of four URL variants and three problem cases identified that there are issues with 12,722 (12.3%) of the 103,539 tested. Surely the problems would extend to an even greater number of articles if we considered additional URL variants or explored other problematic cases. Further work is needed to narrow in on the most problematic scenarios and to quantify the size of the discrepancies that exist.

**Discussion**

The presented results touch upon issues of provenance and resolution that are foundational properties of the web and affect scientometricians and developers alike. Our method presents a first attempt to put a number on the problems and issues around a broader problem underlying scholarly infrastructure and scientometrics. A benefit of this undertaking, in addition to quantifying the size of problems, is the identification of technical issues that exacerbate the situation. In that regard, we echo the recommendations put out by Fenner and Lin (2014) for publishers to optimize pages for Facebook's crawlers with a few additions based on our experiences. We suggest publishers to avoid:

- pages that are not entirely machine readable (i.e., require human intervention)
- too many or even indefinite redirect loops, as Facebook's debugger stops following redirects after the fifth one
- URL injections (e.g., error messages and status codes) that corrupt the results of Facebook's heuristics
- journal pages that are human-readable, but send back an "access not allowed" status
- frequent changes of the directory layout as engagement data collected for previous URLs are not preserved for new Open Graph objects

Furthermore, in some cases the cause of problems can be outside of the publishers control. Wrongly resolved DOIs (e.g., caused by problems at CrossRef's side) can lead to the problem case 3 as well. A complex picture starts to emerge where difficulties arise at several locations such as the original publisher, CrossRef or other linking institutions, and Facebook who conducts the internal mapping of URLs to Open Graph objects. Thus, in order to fully capture Facebook engagements it is necessary to examine data on the levels of publishers and infrastructure and to validate that data by analysing collection of articles, like we have done in this study. The proposed methodology is part of a broader research project with that particular aim. An outline and the progress can be found at the following persistent link of the GitHub repository: DOI 10.5281/zenodo.1338118. At the time of writing, we are currently looking into the differences between public and private engagement on Facebook for various publishers and venues employing different publishing technologies.

**Conclusions**
Working with APIs and URLs on the open web will always present challenges, even within the semi-structured system of scholarly communications. However, there are efforts afoot to tackle these challenges. Today, those researching and using altmetrics are dependent on a handful of data providers (Haustein et al.; 2016). However, new initiatives, like Crossref Event Data are looking to create a place to gather such metrics. As the largest DOI registration agency, they are uniquely positioned to tackle some of the challenges, especially those related to provenance. However, it is incumbent on both researchers and tool builders to continue to investigate the opportunities and limitations of data sources, like we have done here. Our plan is to use what we have learned to build open source software for collecting Facebook metrics that can be used by publishers. The Public Knowledge Project intends to pilot this software to offer this software as a service to journals using its software. This service will consider the challenges we identified, but more research and experiments are needed if the community is going to be able to trust the data collected.